\documentclass[final,3p,times]{elsarticle}

\usepackage{graphicx}
\usepackage{pifont,latexsym,ifthen,rotating,calc,textcase,booktabs,color}
\usepackage{amssymb,amsfonts,amsbsy,amsmath}
\usepackage{mathabx}
\usepackage{float}
\usepackage{upgreek}
\usepackage{subfig}
\usepackage{textcomp}
\usepackage{placeins}
\usepackage{ifthen}
\usepackage{bigstrut}
\usepackage{lineno}

\def\path{\string~/}
\usepackage{bm}
\usepackage{pgfplots}
\usepackage{tikz}
\usepackage{soul}
\usetikzlibrary{shapes, shadows, arrows, positioning,patterns,decorations.pathreplacing}

\biboptions{sort&compress}

\sethlcolor{white}

\begin{document}

\begin{frontmatter}

\title{\hl{Recent progress in CFD modeling of powder flow charging during pneumatic conveying}}

\author[label1,label2]{Holger Grosshans\corref{cor1}}
\ead{holger.grosshans@ptb.de}
\author[label1]{Simon Jantač}

\cortext[cor1]{Corresponding author}

\address[label1]{Physikalisch-Technische Bundesanstalt (PTB), Braunschweig, Germany}
\address[label2]{Otto von Guericke University of Magdeburg, Institute of Apparatus- and Environmental Technology, Magdeburg, Germany}

\begin{abstract}
Thus far, Computational Fluid Dynamics (CFD) simulations fail to predict the electrostatic charging of particle-gas flows reliably.
The lack of a predictive tool leads to powder operations prone to deposits and discharges, making chemical plants unsustainable and prime candidates for explosions.
This paper reviews the rapid progress of numerical models in recent years, their limitations, and outlines future research.
In particular, the discussion includes CFD models for the physics and chemistry of particle electrification.
The condenser model is most popular today in CFD simulations of powder flow electrification but fails to predict most of its features.
New experiments led to advanced models, such as the non-uniform charge model, which resolves the local charge distribution on non-conductive particle surfaces.
Further, models relying on the surface state theory predicted bipolar charging of polydisperse particles made of the same material.
While these models were usually implemented in CFD tools using an Eulerian-Lagrangian strategy, recently Eulerian methods successfully described powder charging.
The Eulerian framework is computationally efficient when handling complete powders;
thus, Eulerian methods can pave the way from academic studies to application, simulating full-scale powder processing units.
Overall, even though CFD models for powder flow charging improved, major hurdles toward a predictive tool remain.
\end{abstract}


\end{frontmatter}


\section{Introduction}

Discharge of electrostatic energy caused numerous catastrophic dust explosions in chemical engineering plants, tremendous economic damage, and the loss of workers' and residents' lives~\citep{Eck03}.
\hl{In the United Kingdom and Germany every 10th day static electricity causes a dust explosion~\mbox{\citep{Glor03}}, in Japanese industry 153 electrostatic accidents are documented over the last 50 years~\mbox{\citep{Osh11}}, and in the United States, 1000 people died or were severely injured for the same reason over 25 years~\mbox{\citep{OshaUS}}.}
Moreover, due to attracting electrostatic forces, particles adhere to the surfaces of pipes, ducts, and other flown-through components.
Growing deposits partly block pipes and increase their surface roughness;
valuable raw material is lost and the system consumes more energy.

One way to control powder flow charging would be to analyze an industrial process through simulations.
Then, based on the results, one could adapt the apparatus' design or choose its operating parameters to limit the generating charge.
However, simulating powder flow charging is exceptionally challenging.
It requires coupling the equations of fluid mechanics (turbulent conveying airflow), surface science (triboelectric charge exchange, adhesion), and electromagnetism (electrostatic attraction of charged particles).
Each of these scientific sub-fields is complex by itself.
Their numerical coupling is yet more difficult.

For some of the sub-processes of powder flow charging, the mathematical equations are not even clear to date.
In particular, the lacking understanding of the physics and chemistry of particle charging explains the limited success of related numerical models.
Particles alter their charge through various physical mechanisms:
through ionized gas or dissipation, but most often through contact with other surfaces.
Models for contact charging usually require heavy tuning of parameters, or the predicted charge differs from experimental measurements by several orders of magnitude.
For these reasons, \hl{numerical simulations} are not mature enough to reliably evaluate the charging of particulates during processing.

\begin{figure}[tb]
\centering
\includegraphics[trim=0cm 0cm 0cm 0cm,clip=true,width=0.50\textwidth]{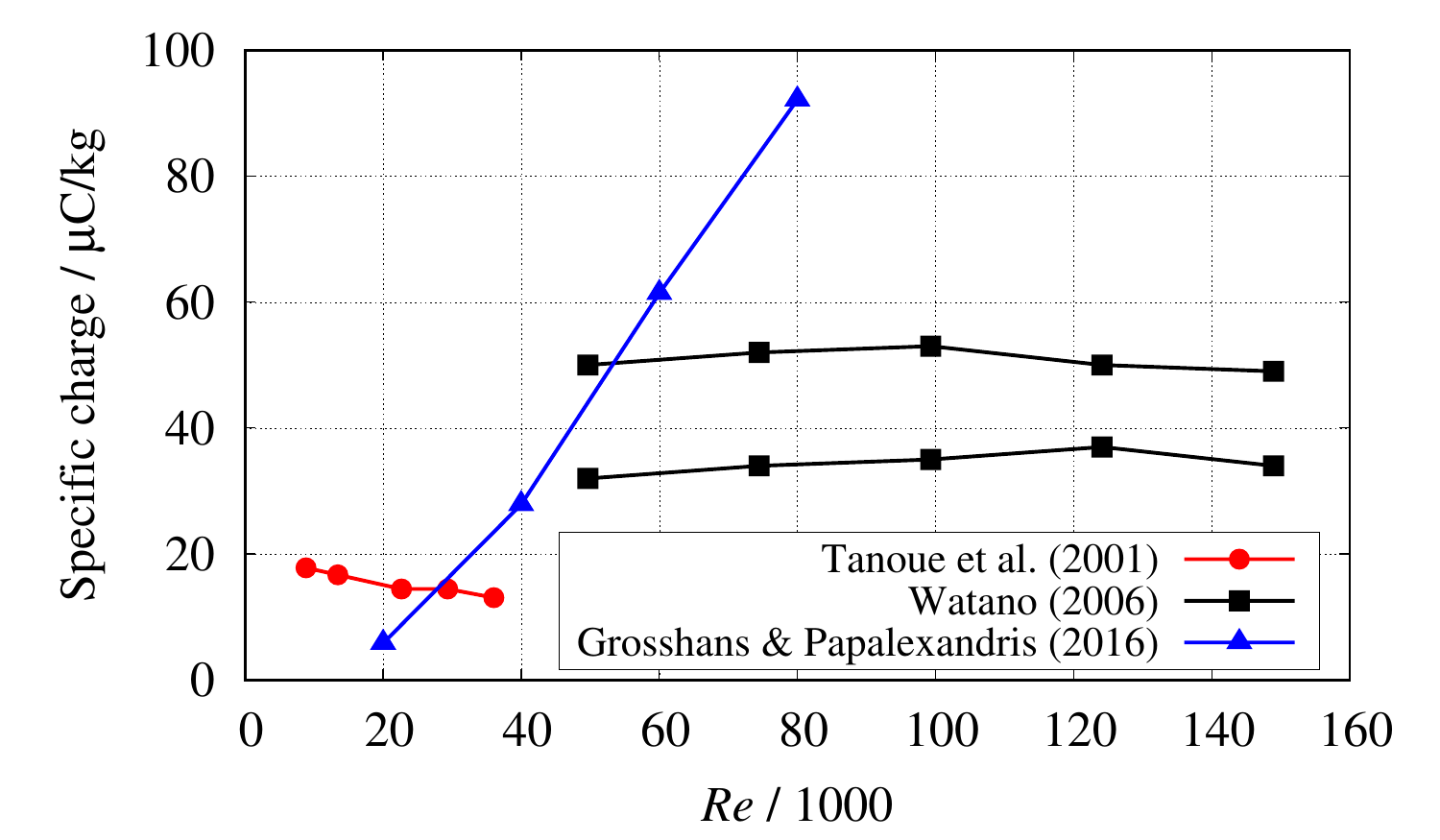}
\caption{\hl{Different CFD methods predict contradicting dependencies of the powder charge on the conveying flow Reynolds number.}}
\label{fig:qre}
\end{figure}

Figure~\ref{fig:qre} compiles CFD predictions of powder charging by three groups using different \hl{numerical methods and models to simulate pneumatic conveying in a duct.}
Having simulated different particle materials and sizes, each group naturally obtains different charge values.
But one would expect all groups to receive the same trend when increasing the Reynolds number, the dominating process parameter;
\hl{namely, either the charge to increase or to decrease.
Each study by itself points in a clear direction, confirming the importance of the Reynolds number.}
However, the results contradict each other:
According to \citet{Wata06}\hl{, who performed DEM (Discrete Element Method)}, the Reynolds number has nearly no influence.
\citet{Tan01}\hl{, using RANS (Reynolds-Averaged Navier-Stokes)}, predict the powder charge to decrease with increasing Reynolds number.
And \hl{the LES (Large Eddy Simulations) of} \citet{Gro16a} suggest the opposite. 
In other words, to reduce the charge on the particles the Reynolds either plays no role, or it must be increased, or it must be decreased.
Thus, the influence of the Reynolds number, as the influence of other process parameters such as powder mass flow rate or Stokes number, remains unclear.
Probably, the contradiction in Fig.~\ref{fig:qre} stems from each group's approach to \hl{turbulence} and multi-physics modelling.

Powder flow electrification is not simply the sum of the charging of the individual particles.
Instead, fluid dynamics, electrostatics, and triboelectricity give rise to complex intertwined interactions, e.g.:
\begin{itemize}
\itemsep1mm
\item The dynamics of a particle-laden flow determine the frequency and severeness of particle/surface and particle/particle contacts and, thus, the charge accumulation of powder~\citep{Gro17a,Jin17}.
\item The charge exchange during one contact does not only depend on the charge carried by the particle itself but also on the electrostatic field generated by all other particle charges and induced charges on surfaces~\citep{Mat95c,Mat97}.
\item The electrostatic field significantly changes the powder flow pattern through electric forces and, thus, alters the dynamics of subsequent contacts \citep{Dho91}.
\end{itemize}
Due to the emerging electrostatic field, these interactions cause perplexing phenomena, such as particles moving counter to the main gas flow~\citep{Myl87}.
In other words, only having a correct particle charging model is not enough for a correct prediction of powder charging.
In essence, the hazard of electrostatic charging to process safety must be evaluated at a powder flow level.

This paper reviews the state-of-the-art, limitations, and progress in recent years of the numerical modeling of electrostatic charging of particle-gas flows.
Out of all industrial powder operations, pneumatic conveying, due to the high flow velocities, leads by far to the highest charge levels~\citep{Kli18}.
Often the transport in the conveying line toward the actual apparatus generates most of the powder charge.
Also, pneumatic conveying resembles a generic wall-bounded flow, which is part of most other powder processing operations.
Therefore, this review focuses on simulations of pneumatic conveying.
\hl{It is noted that the operation that generates the most charge is usually separated from the location that is most prone to discharges and explosions. 
Discharge happens where the generated charge excessively accumulates, for example, in silos of filters.}

Nevertheless, the research questions in pneumatic powder conveying are often similar to those of closely related fields, and their model development stimulates each other.
Thus, this review also touches the progress of other fields.
Moreover, this review summarizes advances in simulations;
purely experimental studies are only included if they directly led to a model.
Otherwise, the reader is referred to the reviews of \citet{Lacks19} on general triboelectricity, \citet{Chow21} on single-particle charging models, \citet{Mat10} on experimental electrostatics, \citet{Wei15} on experiments and liquid charging, \citet{Mehr17} on charging in fluidized beds, and \citet{Wong15} on charging in pharmaceutics.
\hl{Furthermore, several recent articles reviewed the CFD modeling of pneumatic conveyors omitting the effects of electrostatic charging \mbox{\citep{Wang17,Kuang20,Shi22}}.}

This paper is organized as follows:
Sections~2 to~4 present the available numerical concepts to model the flow of charged powder in pneumatic conveying.
More specifically, Section~2 gives an overview of the methods to simulate the carrier gas phase.
Section~3 outlines the different methods to simulate the dynamics of powder, including approaches to compute the electric field and the electrostatic forces on the particles.
Section~4 summarizes the models of triboelectric charging on a single particle level.
The final section concludes and gives the authors' opinion on the future perspectives of the field.

\section{Modeling the turbulent carrier gas flow}
\label{sec:gas}

Given that particles collect most of their charge during contacts, and contacts are driven by aerodynamic forces, the simulation of the carrier gas flow plays a paramount role in powder charging.
The gas flow in powder processing facilities is described by the Navier-Stokes equations.
That means by the mass and momentum balance of incompressible Newtonian fluids in the Eulerian framework,
\begin{subequations}
\begin{equation}
\label{eq:mass}
\nabla \cdot {\bm u} = 0
\end{equation}
\begin{equation}
\label{eq:mom}
\frac{\partial {\bm u}}{\partial t} + ({\bm u} \cdot \nabla) {\bm u}
= - \frac{1}{\rho} \nabla p  + \nu \nabla^2 {\bm u} + {\bm f}_\mathrm{s}\, ,
\end{equation}
\end{subequations}
where ${\bm u}$ denotes the fluid's velocity, $p$ its pressure, $\rho$ its density, $\nu$ its kinematic viscosity, and $t$ the temporal coordinate.
The source term ${\bm f}_\mathrm{s}$ accounts for the momentum transfer from the particles to the carrier fluid.

Most of the time, pneumatic conveyors operate at high Reynolds numbers, which means fully turbulent mode.
The most exact method to simulate turbulence, \textit{direct numerical simulation}~(DNS), resolves all length- and time-scales of fluid motion.
Resolving all scales of high Reynolds number flows requires a fine numerical grid and a small time-step, resulting in a high computational effort.
Therefore, turbulence is usually modeled instead of resolved when simulating pneumatic powder conveying.

In early computations of powder charging, not even the mean flow was solved but approximated by an analytical velocity profile.
Afterward, the first simulations appeared using the RANS approach~\citep{Kol89,Tan99,Tan01}.
In RANS, Eqs.~(\ref{eq:mass}) and~(\ref{eq:mom}) are temporally or ensemble-averaged.
Due to the averaging, new unclosed terms arise, the so-called Reynolds stresses.
\hl{A large number of available turbulence models approximate these terms.}

RANS is reasonable when only mean quantities are of interest rather than turbulent fluctuations.
However, turbulence drives the impact frequency, velocity, and angle of particles on the conveying duct's walls.
Thus, the turbulence model's deficiency directly impairs the prediction of powder charging by RANS simulations.

For several years, LES of powder flow charging has been feasible~\citep{Kor14,Gro16a}.
LES computes the filtered governing Eqs.~(\ref{eq:mass}) and~(\ref{eq:mom});
the turbulent motions larger than the filter size are resolved on the grid.
Thus, the computational effort of LES is much higher compared to RANS.

Similar to RANS, in LES new terms corresponding to the small (subfilter) scales appear through the filter operation, \hl{which need closure through turbulence models.
Popular RANS and LES turbulence models, their application range, and their limitations are discussed in textbooks \mbox{\citep{Pope00}}.
However, despite the well-known shortcomings of these models, the LES of charged powder flows is reliable under two conditions:
First, if a sufficiently fine grid resolves most of the turbulence energy spectrum.
And second, if the ratio of the characteristic time of a particle (or droplet), $\tau_\mathrm{p}$, to the characteristic time of the flow, $\tau_\mathrm{f}$, is high.
This ratio is the particle's Stokes number,}
\begin{equation}
S\!t=\dfrac{\tau_\mathrm{p}}{\tau_\mathrm{f}},
\end{equation}
that determines the dynamics of the turbulence-particle interaction.
For particles of a high Stokes number, inertial forces act as a high-pass filter.
Their trajectories are influenced by large-scale but not by small-scale turbulence.
Thus, the requirement for the grid resolution relaxes when simulating the charging of high Stokes number particles.

\begin{figure} [tb]
\centering
\subfloat[Pipe cross-section]{
\includegraphics[trim=16.4cm 0cm 0.0cm 0.0cm,clip=true,width=0.12\textwidth]{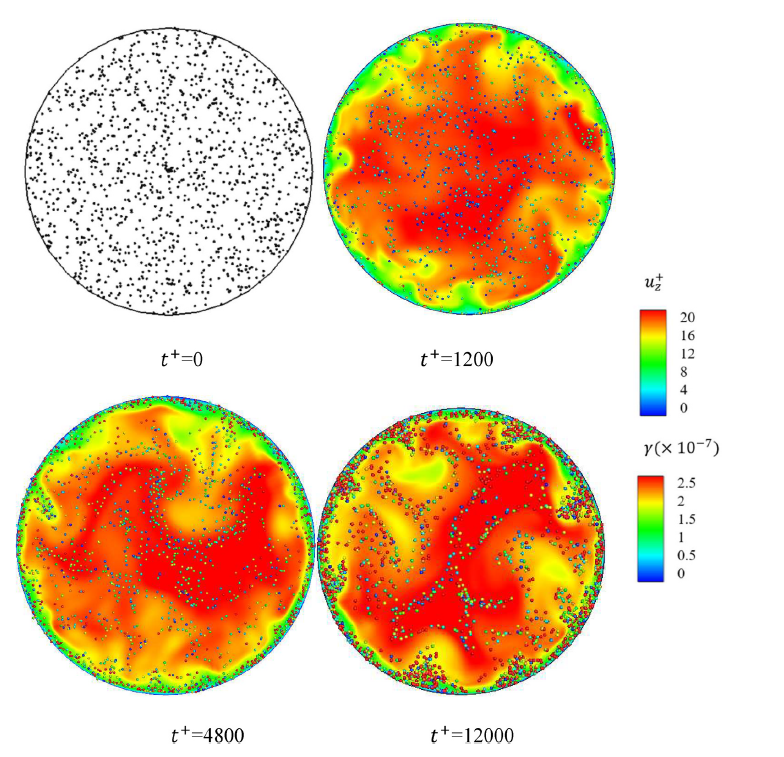}
\includegraphics[trim=8.4cm 0cm 4.5cm 0.2cm,clip=true,width=0.17\textwidth]{FIG5_LargeEDDy}
\label{fig:LESa}
\qquad
\qquad
\qquad
}
\subfloat[Side view, near-wall region]{\includegraphics[trim=0cm 0cm 0cm 0cm,clip=true,width=0.5\textwidth]{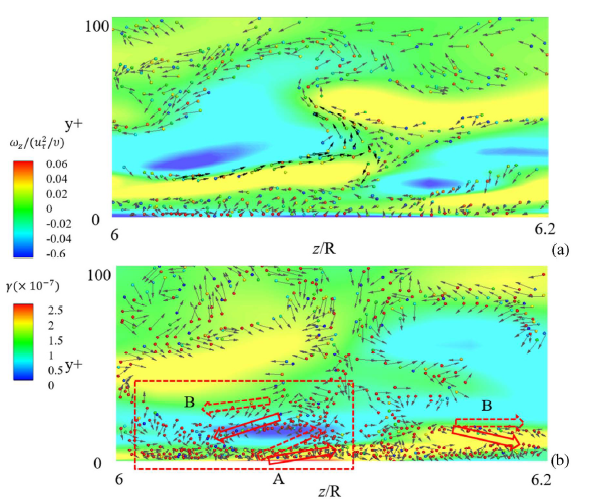}
\label{fig:LESb}
}
\caption{Pipe flow, $Re=$~44\,000, $S\!t=$~142.2.
Normalized particle charge ($\gamma$), streamwise gas velocity ($u_z^+$), and gas vorticity ($\omega_z$) at two time instances: $t^+=$~1200 (top) and $t^+=$~12\,000 (bottom).
In (b), the solid red arrows show the particle velocity vector.
For comparison, the dashed arrows give the particles' direction without electrostatic forces~\citep{Li21} (adapted with permission).
}
\label{fig:LES}
\end{figure}

The LES of \citet{Li21} of pipe flows of a Reynolds number as high as 44\,000 revealed the complex interplay between large-scale turbulence and the particles' triboelectrification.
They compared three $S\!t$ numbers, 3.9, 35.6, and~142.2.
For $S\!t=$~142.2, Fig.~\ref{fig:LES} shows the gas flow and the particles' location and charge.
Figure~\ref{fig:LESa} depicts the cross-section of the flow at two time instances, $t^+=$~1200, when the charge begins to build up, and $t^+=$~12\,000, when the particles approach their saturation.
The particles tend to move toward the pipe's wall due to turbophoretic drift, caused by the inhomogeneity of wall-bounded turbulence.
Also, the particles' charge distributes inhomogeneously.
Those particles within the boundary layer charge faster than those close to the pipe's centreline because the boundary layer is where the particles collide with the wall and receive charge.

For the same two time instances, Fig.~\ref{fig:LESb} shows the influence of the streamwise vortex structure on the particles in the boundary layer.
At the earlier time, when the particles carry little charge, they gather around the periphery of the vortex structures.
Thus, the flow momentum dominates particle transport.
Later (bottom), when the particles' charge increases, they penetrate the interior of the vortex structures.
At the position marked in the figure with "A", the dashed arrows indicate that electrostatic forces delay the ejection of particles from the wall toward the vortex.
Conversely, at position "B", electrostatics accelerates the sweep of particles along the vortex periphery to the wall.
Both explain the increase of the particle concentration in the boundary layer under the influence of electrostatic forces.

\hl{The same group used LES to simulate the behavior of charged powder in a 90-degree bend at Reynolds numbers of 34\,000, 40\,000, and 58\,000~\mbox{\citep{Zhao21}}.
The particles' charge was assumed constant.
According to their LES, the particles
aggregate near the wall, resulting in lower impact angles and, thus, a decreased erosion rate.
This effect is strongest for low airflow rates where electrostatic forces dominate.
However, LES has also shown that low airflow rates cannot limit charging, as shown by \mbox{\citet{Cer19}}.
Using a modified capacitor model for inter-particle and particle-wall charge transfer, they found that an increased velocity results in fewer particle-wall collisions.
But when such collisions occur, more charge is transferred because the contact area is larger.
Moreover, low air velocity can lead to the settling of particles.
Then, the particles roll on the conveying line's bottom wall and charge extensively.}

\begin{figure}[tb]
\centering
\subfloat[]{\includegraphics[trim=0mm -15mm 120mm 0mm,clip=true,width=.48\textwidth]{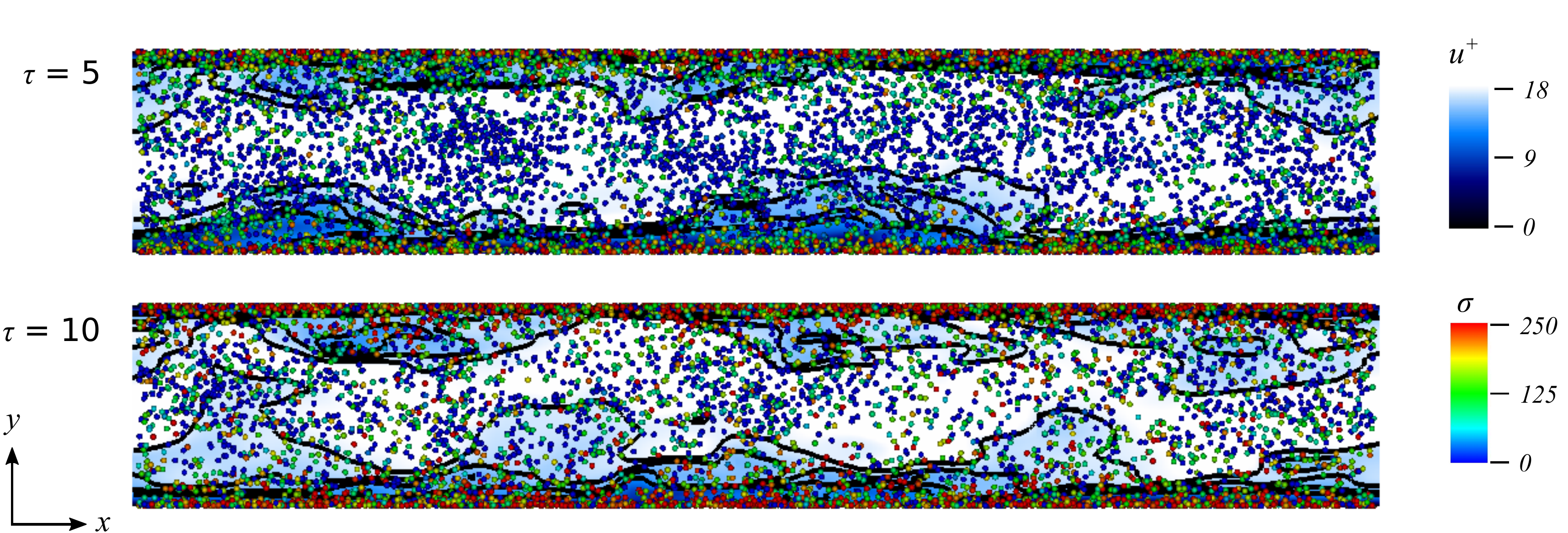}%
\label{fig:plates}%
}
\quad
\subfloat[]{
\begin{tikzpicture}[scale=1]
\node at (0,0) {\includegraphics[trim=0cm 0cm 0cm 0cm,clip=true,width=.45\textwidth]{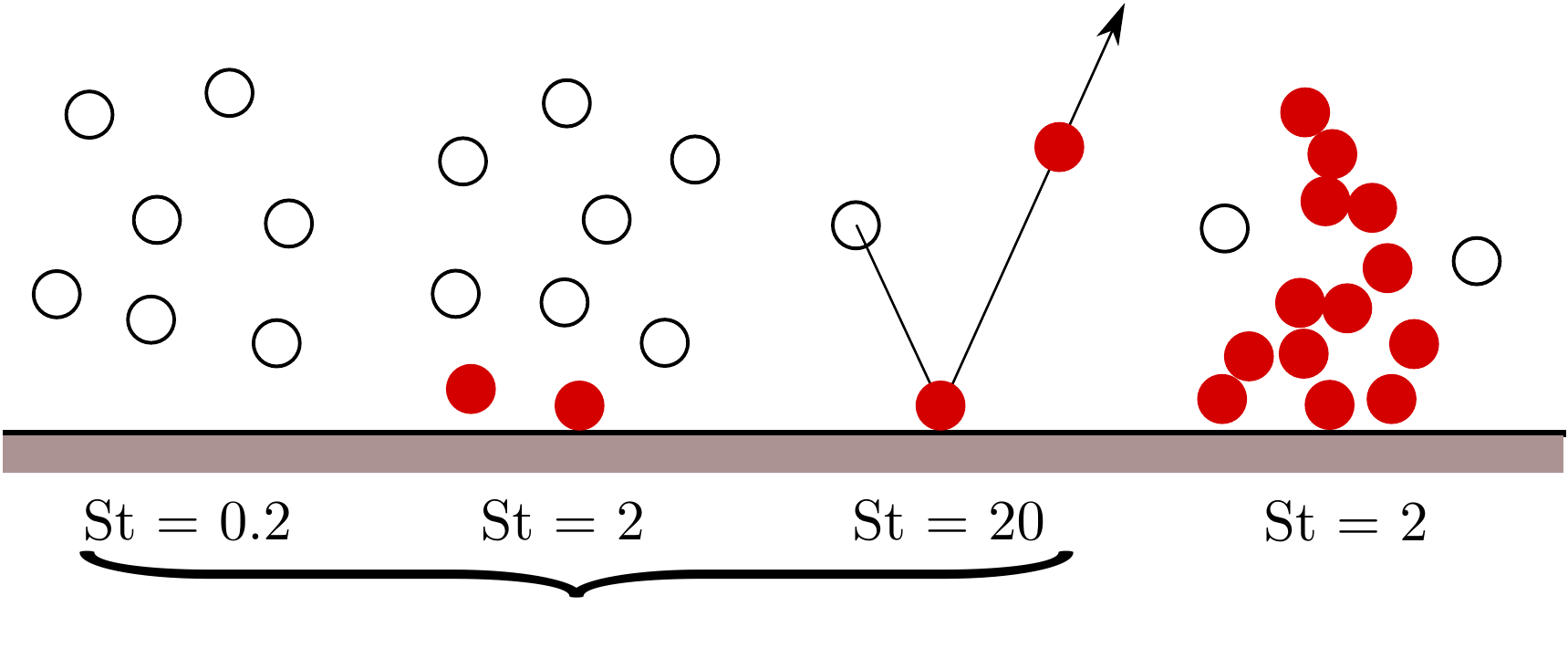}};
\draw [->,>=latex,ultra thick] (-3.0,2.0) node [above right,black,font=\footnotesize,xshift=0] {$u_\mathrm{gas}$} -- (-2.0,2.0);
\draw [->,>=latex,ultra thick,blue] (.8,2.0) node [above,align=center,black,font=\footnotesize,xshift=-15] {particle-bound\\charge transport} -- (.8,1.3);
\draw [->,>=latex,ultra thick,blue] (2.9,2.0) node [above,align=center,black,font=\footnotesize] {inter-particle\\charge diffusion} -- (2.9,1.3);
\node at (-.8,-1.8) [above,font=\footnotesize] {low $\phi$};
\node at (3.0,-1.8) [above,font=\footnotesize] {high $\phi$};
\end{tikzpicture}%
\label{fig:mech}}
\caption{DNS of powder electrification in a channel flow depending on the Stokes number ($S\!t$) and particle volume fraction ($\phi$).
(a) is for $S\!t=$~20, the colors indicate the particles' charge~\citep{Gro17a,Gro18c}
(adapted with permission).}
\end{figure}
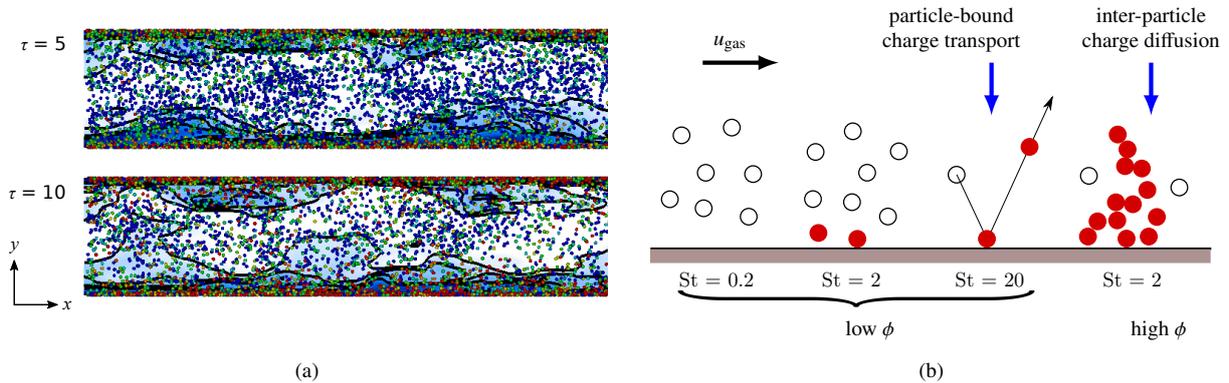

Only recently, the first DNS of electrifying powder flow was achieved~\citep{Gro17a}.
However, DNS can not simulate complete industrial unit operations.
Instead, it is limited to generic domains and low Reynolds numbers, such as the channel flow of a friction Reynolds number of~360 in Fig.~\ref{fig:mech}.
These DNS revealed, at a previously unknown level of detail, the small-scale mechanisms that determine the powder charging rate.
More precisely, the mechanisms sketched in Fig.~\ref{fig:mech} dominate the charge transfer from the channel walls to and within the powder flow:
\textit{particle-bound charge transport} for highly inertial particles and \textit{inter-particle charge diffusion} for low inertial particles in case of high particle volume fractions.
Identifying these mechanisms implies the possibility of controlling the electrification of powder flows by imposing flow conditions that purposely trigger these mechanisms.

\section{Modeling electrostatically charged powder flow}
\label{sec:powder}

Contrary to the carrier gas, which is continuous, powder forms a dispersed phase consisting of abundant particles.
The number of particles and their solid/gas interface area restricts the choice of the numerical method.
Numerical multiphase methods that resolve the phase interface on the grid are computationally too expensive.
Instead, pneumatic conveying is usually modeled by the Eulerian-Lagrangian or the Eulerian-Eulerian approach.
In both approaches, the carrier gas is described in the Eulerian framework, as discussed in Sec.~\ref{sec:gas}.
The particulate phase is either described in the Lagrangian or the Eulerian framework. 
That means the particles are either tracked individually or modeled as a continuum.

Sub-section~\ref{sec:forces} reviews the forces on charged particles.
The following Sub-sections~\ref{sec:lagrange} and~\ref{sec:euler} elaborate on the challenges of Lagrangian and Eulerian formulations.

\subsection{Forces on a charged particle}
\label{sec:forces}

\begin{figure} [tb]
\centering
\includegraphics[trim=0cm 0cm 0cm 0cm,clip=true,width=0.4\textwidth]{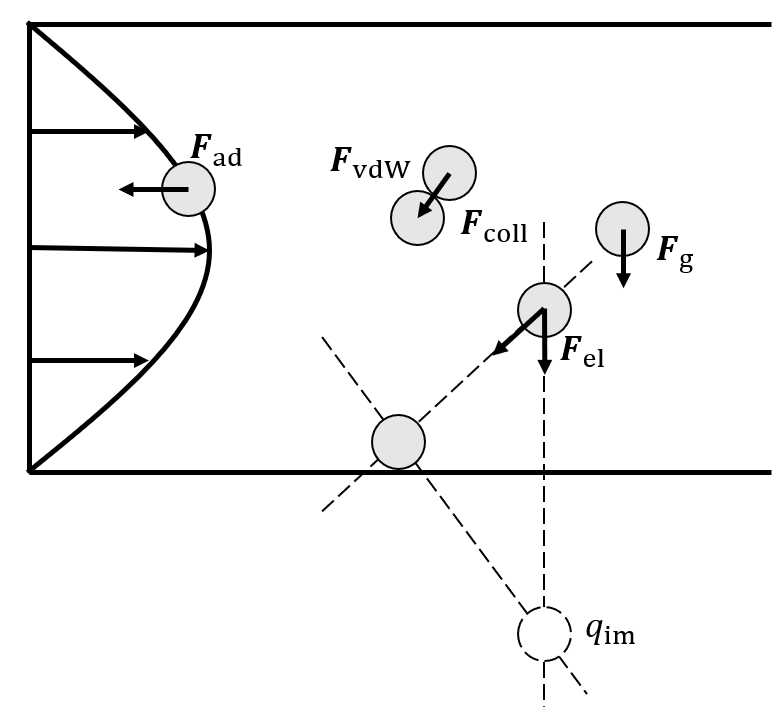}
\caption{
Gravitational (${\bm F}_{\mathrm{g}}$), collisional (${\bm F}_{\mathrm{coll}}$), drag (${\bm F}_{\mathrm{ad}}$), electrostatic (${\bm F}_{\mathrm{el}}$), and van der Waals (${\bm F}_{\mathrm{vdW}}$) forces affect charged particles in pneumatic conveyors.
Only electrostatic forces act over long distances.
The particles' image charges on conductive surfaces, $q_\mathrm{im}$, contribute to the electric field~\citep{Kor14} 
(adapted with permission).}
\label{fig:LagrangianSchematics}
\end{figure}

The trajectories of charged particles in a gas flow are affected by gravitational, collisional, drag, van der Waals, and electrostatic forces.
As sketched in Fig.~\ref{fig:LagrangianSchematics}, contrary to the other forces, the electrostatic forces reach over long distances.
The selection of forces included in the model depends on the simulated conveying system:
for vertical conveying of high Stokes number particles, the particle dynamics with and without gravity are nearly identical~\citep{Marc07};
thus, the gravitation can be neglected.
For horizontal conveying of low Stokes number particles, gravity determines the particles' trajectories and, hence, their charging.
Therefore, gravitation is considered in all simulations of horizontal conveying.

The collisional force accounts for inter-particle and particle-wall collisions.
Collisions between particles require the comparison of particle pairs, which is computationally expensive, as discussed in detail in the following sub-section.
Therefore, inter-particle collisions are neglected whenever possible.
During dilute conveying, particles seldom collide with each other~\citep{Elg94}.
Therefore, inter-particle collisions are usually only modeled when simulating dense conveying.

Due to the high flow velocities, all pneumatic conveying simulations contain the aerodynamic drag on the particles.
Classical correlations for the drag coefficient were derived for idealized conditions, namely for isolated, spherical particles exposed to an undisturbed airflow.
These idealizations generally do not hold for pneumatic conveying.
Many new drag correlations were proposed in recent years, reflecting non-spherical particles~\citep{Wac12}, shear flow due to the pipe's walls~\citep{Zeng09}, or the disturbance of the flow by nearby particles~\citep{Krav19,Tang14}.
However, the dynamics of charged particles are different from uncharged ones, and so is their drag. 
The drag correlation for charged particles has not been researched yet, except in the thesis of \citet{Ozl22}.
Given that the near-wall dynamics of particles determine their charging during pneumatic conveying, choosing a suitable drag correlation is decisive for predicting powder flow charging.

Other aerodynamic forces (besides drag), summed up by the Basset-Boussinesq-Oseen~(BBO) equation~\citep{Max83}, include the virtual mass, Faxen, Saffmann, and Basset history forces.
Except for the Saffmann lift forces, these are typically neglected for pneumatic transport.

Van der Waals forces can be stronger than gravitational forces if the particles are small~\citep{Tom09}.
During pneumatic conveying, van der Waals forces play no role for airborne particles. 
They act only during a minuscule duration when the distance between two particles or a particle and a wall is of the nanometer order;
during this small duration, the particle's momentum changes marginally.
Nevertheless, van der Waals forces can lead to dust deposits on the surfaces of pipes or other components.
Thus, for the prediction of deposits, van der Waals forces need to be considered~\citep{Gro19e}.

Finally, the electrostatic force acting on a particle that carries the charge $Q$,
\begin{equation}
\label{eq:fel}
{\bm F}_{\mathrm{el}} = Q \, {\bm E} \, ,
\end{equation}
can dominate the dynamics of particles in pneumatic conveyors.
The electric field strength, ${\bm E}$, is given by Gauss' law,
\begin{equation}
\label{eq:gauss}
\nabla \cdot {\bm E} = \dfrac{\rho_{\mathrm{el}}}{\varepsilon} \, ,
\end{equation}
where $\varepsilon$ is the electrical permittivity and the electric charge density, $\rho_{\mathrm{el}}$, reflects the charge carried by all particles in the system.
Gauss' law involves only $O(N)$ operations and is, therefore, fast to solve.
However, an extremely fine grid is required to resolve the electric field gradient caused by charged particles in close proximity.

Assuming the charge of each particle is located at its centre point, a mathematical equivalent formulation to Eq.~(\ref{eq:gauss}) is Coulomb's law,
\begin{equation}
\label{eq:coulomb}
{\bm E}_m = \sum\limits_{n=1,n\neq m}^N \dfrac{Q_n \, {\bm z}_{n,m}}{4 \, \pi \, \varepsilon \, |{\bm z}_{n,m}|^{3}} \, .
\end{equation}
Herein, ${\bm E}_m$ is the electric field at the position of particle $m$, $N$ the number of all particles in the system, and ${\bm z}_{n,m}$ a vector pointing from the center of particle $n$ to the center of particle $m$.

Equation~(\ref{eq:coulomb}) contains only Lagrangian variables and, therefore, requires no grid to solve.
A drawback compared to Eq.~(\ref{eq:gauss}) is that it involves comparisons of particle pairs, thus, $O(N^2)$ operations.

Similar solutions to this problem were independently proposed by \citet{Kol16} and \citet{Gro17e}, combining the numerical advantages of Gauss' and Coulomb's law.
More specifically, their hybrid approaches superimpose the far-field interactions computed with Eq.~(\ref{eq:gauss}) and the coulombic interactions between the particle and its neighbors.
This approach is both fast and accurate and generally recommended for future simulations.
In particular, it is more suitable for wall-bounded flows than the Ewald summation or the P$^3$M (Particle-Particle-Particle-Mesh) method~\citep{Yao18}.
\hl{The Ewald summation calculates long-range interactions in systems with periodic boundaries.
The P$^3$M method is based on the Ewald summation for long-range interactions (particle-mesh);
thus, both methods are cumbersome to implement in complex geometries.
Below a cutoff distance, P$^3$M calculates interactions pairwise (particle-particle).
This allows coarser grids while keeping the accuracy high.}

Nevertheless, the point charge assumption impedes the prediction of particle dynamics resulting from inhomogeneous charge distribution on the particles' surface.
For example, the attraction of particles of the same polarity due to induced charges~\citep{Qin16} cannot be captured.
For fluidized beds, \citet{Kol18b} recently included particle polarization due to surrounding charges.
The development of advanced numerical models reflecting the surface charge distribution is expected to boost the accuracy of future pneumatic conveying simulations.

Besides the charge carried by the particles, their image charge on grounded and conducting surfaces (see Fig.~\ref{fig:LagrangianSchematics}) contributes to the electric field.
If Eq.~(\ref{eq:gauss}) is solved, the image charges go in the boundary conditions at the surface:
zero-gradient for ${\bm E}$, zero for $\rho_\mathrm{el}$, or zero for the electric potential; these three formulations are equivalent.
However, if Eq.~(\ref{eq:gauss}) is not solved, if electrostatic forces are calculated only by coulombic interactions (Eq.~(\ref{eq:coulomb})), then the method of images is convenient.
In the method of images, a fictitious image charge is added at the same distance but the opposite side of the surface.
This image charge holds the same magnitude but opposite sign than the charge of the originating particle.
At the surface, the potential of the charged particle and its image yields zero, showing the equivalence of the method of images to the boundary conditions of Eq.~(\ref{eq:gauss}).

However, the charged particle not only induces the image at the lower surface sketched in Fig.~\ref{fig:LagrangianSchematics} but also at the upper surface. 
If another fictitious charge is added for that image, the potential at the surfaces is not zero anymore.
The boundary condition is corrected by adding the images of the images, which again need images, leading to an infinite series of images.
\citet{Kor14} truncated the series after two terms, which caused a maximum error of 5\% for a particle close to a surface.

\subsection{Lagrangian}
\label{sec:lagrange}

Most simulations of pneumatic conveying \hl{and, thus, most papers discussed in this review} describe the flow and charging of particles in the Lagrangian framework.
In the Lagrangian framework, each particle is treated individually as a point-mass whose motion is computed as 
$m_\mathrm{p} \, {\bm \dot{u}}_\mathrm{p} = \sum {\bm F}$,
where ${\bm u}_\mathrm{p}$ is the velocity and $m_\mathrm{p}$ the mass of the particle.
The term on the right-hand sums up all external forces acting on the particle, as elaborated above.

The advantage of the Lagrangian approach is that there is no limitation on $S\!t$ and polydispersity can be handled more easily compared to the Eulerian approach.
However, the particles must be smaller than the characteristic flow scale.
Further, the numerical coupling of particles and gas flow, i.e., Lagrangian and Eulerian framework, poses a challenge.

The computational effort of the Lagrangian approach scales with the number of particles, $N$.
Some sub-models scale linearly with $N$.
Others, such as collisions between particles, require the comparison of particle pairs.
The computational effort of comparing particle pairs scales by $O(N^2)$.
Advanced algorithms reduce the cost, for example, Fast Multipole Methods (FMM)~\citep{Rok90} to $O(N\log N)$.
Nevertheless, operations that require evaluating particle pairs remain elaborative.
Especially for pneumatic conveying systems, which consist of millions of particles, these operations can easily inflate the overall computational time.
Therefore, the models describing Lagrangian particles must be carefully chosen to optimize the equation system's accuracy and efficiency.

Further, the Lagrangian framework is limited to studying the transport through one pipe instead of a complete pneumatic system and for dilute or pulsed conveying where the particle number is low. 
Or, for academic research, looking at fundamental charging methods in only a section of the complete pipe.
The Lagrangian approach plays out its strength for fundamental research, namely its ability to resolve individual particle trajectories.

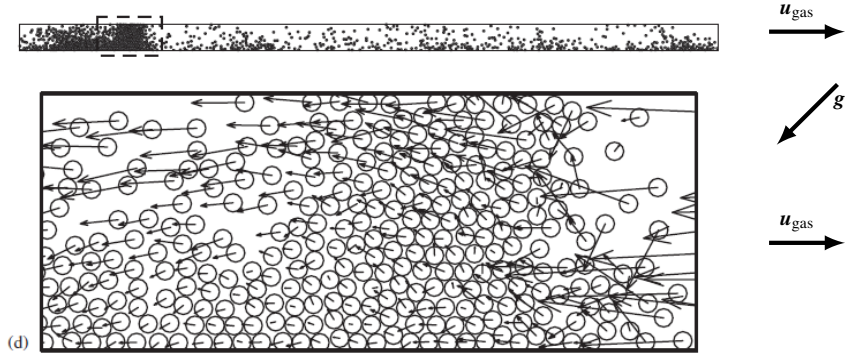
\begin{figure} [tb]
\centering
\begin{tikzpicture}[scale=1]
\node at (0,0) {\includegraphics[trim=0cm 0cm 0cm 11.7cm,clip=true,width=0.65\textwidth]{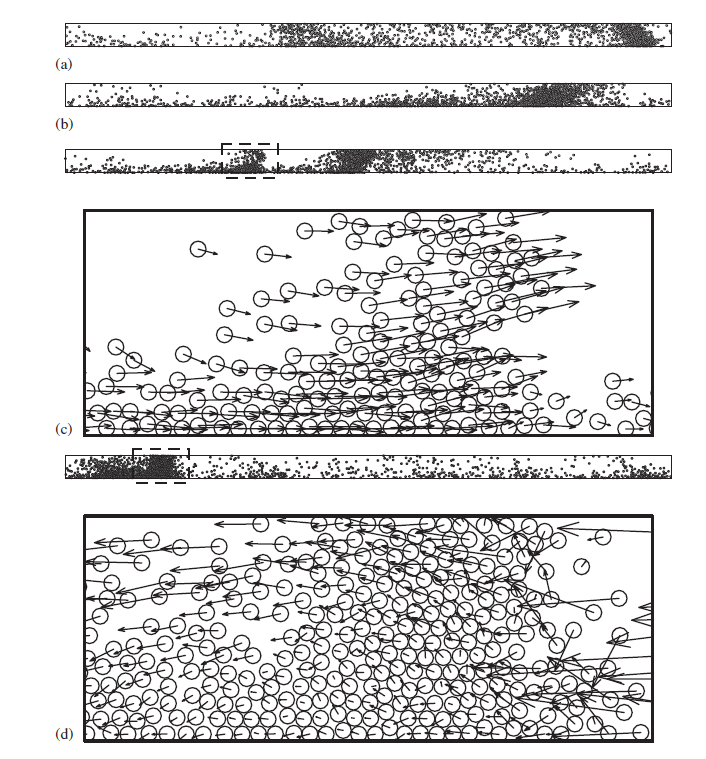}};
\draw [->,>=latex,ultra thick] (5.5,2.1) node [above right,font=\footnotesize,xshift=0] {${\bm u}_\mathrm{gas}$} -- (6.5,2.1);
\draw [->,>=latex,ultra thick] (5.5,-0.7) node [above right,font=\footnotesize,xshift=0] {${\bm u}_\mathrm{gas}$} -- (6.5,-0.7);
\draw [->,>=latex,ultra thick] (6.4,1.4) node [below,font=\footnotesize,xshift=0.5] {${\bm g}$} -- (5.6,0.6);
\end{tikzpicture}
\caption{Reverse flow of highly charged particles during conveying through a 45$^\circ$ inclined pipe~\citep{Chu06}
(adapted with permission).}
\label{fig:CFD-DEM_Simple}
\end{figure}

Through Eulerian-Lagrangian simulations, \citet{Chu06} showed that charged particles can move in reverse direction, counter to the main flow.
Additional to the linear, they solved the particles' angular momentum 
They simulated upward pneumatic conveying through a 45$^\circ$ inclined pipe with a gas velocity of 5~m/s.
Uncharged particles travelled in the moving dunes regime, clusters sliding along the bottom wall of the pipe.
In the moving dune regime, most particles move in the direction of the airflow.
Then, they assigned constant and unipolar electrostatic charges to the particles.
When assigned small charges, few particles started to move in reverse direction along the bottom wall.
Being highly charged, as can be seen in Fig.~\ref{fig:CFD-DEM_Simple}, most particles move in opposite direction of the gas flow.
They attributed the reverse motion to deceleration due to adherence on the bottom wall and resulting frictional forces.
In their configuration, gravitation accelerated particles backwards.
However, backflow without gravitational assistance, purely driven by electrostatic forces, has only been observed experimentally~\citep{Myl87}.
The numerical reconstruction of his flow phenomenon is yet lacking.

\hl{Afterward, \mbox{\citet{Lim12}} extended their previous study~\mbox{\citep{Chu06}} to vertical and horizontal pipes.
Therein, a layer of charged particles formed on the walls at low gas velocities.
The induced charge between the charged particles and the grounded conveyor wall resulted in ring flow patterns.

\mbox{\citet{Gro16b}} identified the statistically significant process parameters that influence charging in vertical pipes by the Lagrangian approach.
In contrast to \mbox{\citet{Lim12}}, the charging process was modeled.
Thus, the simulations captured the nonequilibrium phase.
According to their results, the most significant parameters were the particle relative permittivity, elasticity, resistivity, and gas velocity.
On the other hand, the particle mass flow rate, radius, and pipe elasticity had only a minor effect on the resulting charge.}

\subsection{Eulerian}
\label{sec:euler}

The Eulerian-Lagrangian approach especially suits numerical studies of laboratory-scaled systems.
But even the expense of $O(N)$ operations limits the number of particles that can be computed simultaneously.
Contrary, the Eulerian framework opens the possibility of handling full technical flows consisting of many particles.
The Eulerian description treats the powder as a continuum with averaged properties in each computational cell.

\begin{figure}
\vspace{-\ht\strutbox}
\centering
\includegraphics[trim=0mm 0mm 0mm 0mm,clip=true,width=0.8\textwidth]{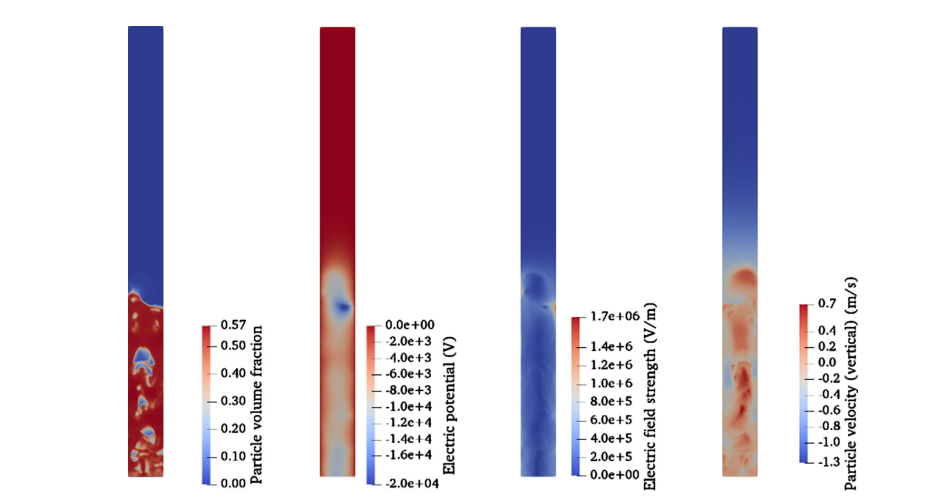}
\caption[]{\hl{Euler-Euler simulations of the triboelectric charging of insulating particles in a fluidized bed~\mbox{\citep{Ray18}}.
The results agree well with experiments
(reprinted with permission).}}
\label{fig:Ray19}
\end{figure}

While the Eulerian-Eulerian approach is popular for general powder flow simulations, only recently, a few studies appeared where it was employed for the charge generation of particle-laden flows.
\hl{Since fluidized beds contain many times more particles than pneumatic conveyors, research on electrification in fluidized beds mainly drove Eulerian formulations forward.
\mbox{\citet{Rokk10,Rokk13}} developed the first transient two-fluid models that involved electrostatic interactions. 
They successfully predicted the radial segregation of charged particles in the fluidized beds;
however, their approach was limited to particles of constant charge.
\mbox{\citet{Kol18}} derived a two-fluid model with a charge transfer mechanism whose form is similar to the heat transfer term in granular flows.
An additional second term accounted for the electric field effect on charged particles.
For high particle volumetric fractions, their model yielded similar results to DEM, making it suitable for dense flows.
\mbox{\citet{Ray18}} extended the approach of \mbox{\citet{Kol18}} by a correlation of the particle charge and velocity.
Doing so broadened the validity of two-fluid models to lower volumetric fractions and stronger electric fields.
Their results in Fig.~{\ref{fig:Ray19}} compared well to experiments of polyethylene particles in bubbling fluidized beds.
\mbox{\citet{Mont20}} derived less restrictive closures for charge transport equations, especially for charge velocity correlations.
Further, they split the charge dispersion coefficient into a collisional one that dominates dense regimes and a kinetic one that dominates dilute flows.
}

\hl{Whereas the mentioned works are limited to mono-disperse particle size distributions, \mbox{\citet{Ray20}} expanded their earlier model to bi-disperse granular flows.
In their model, each particle size has its phase, thus, making simulations of strongly poly-disperse systems cumbersome.}
Finally, \citet{Gro20h} presented a description for the transport of charged poly-disperse powder in the Eulerian framework using the direct quadrature method of moments~(DQMOM)~\citep{Mar05}\hl{, which does not require unique phases for each particle size.}

All these Eulerian formulations are steps toward simulating the charge build-up in technical flow facilities.
Nevertheless, the accuracy of these models lacks way behind Lagrangian formulations.

\section{Particle charging models}

All methods to simulate pneumatic powder conveying discussed in the previous Section assume the particles to be smaller than the cells of the computational mesh.
In other words, the numerical grid does not resolve the gas-solid interfaces.
Thus, all physical processes on the particles' surface need to be modeled explicitly.
In general, those processes include, for example, aerodynamic drag, heat and mass transfer, collisions, phase change, adhesion, and chemical reactions. 
However, most current papers reduce those processes to aerodynamic drag and interactions among particles and conveyor walls.
The electric field created by charged particles is often omitted from the simulations, or overly simplified assumptions are used, even though the electric field can significantly alter or hinder the operation of pneumatic systems.
To obtain a computationally efficient model suitable for CFD simulations, complex physical mechanisms need to be simplified.
Usually, the uncertainty of the particle models leads the error of the overall simulation model.

The implementation in a CFD approach requires the model to be accurate, computationally efficient to handle a vast amount of particles, able to predict charge transfer based on the data available in a CFD framework, and valid for conditions relevant to technical flows.
These requirements impede the usage of detailed theoretical approaches, such as quantum mechanical or atomistic calculations~\citep{Fu17}.
Because of the limits of the models for particle charging in pneumatic conveyors, this Section extends to neighboring fields such as fluidized beds, contact electrification on the sub-particle level, or atmospheric science~\citep{Har16}.
\hl{For reviews dedicated to triboelectric charging models, the reader is referred to the papers cited in the introduction.}

\subsection{Charge limitation}

\begin{figure}[b]
\vspace{-\ht\strutbox}
\centering
\includegraphics[trim=0mm 0mm 0mm 0mm,clip=true,width=0.42\textwidth]{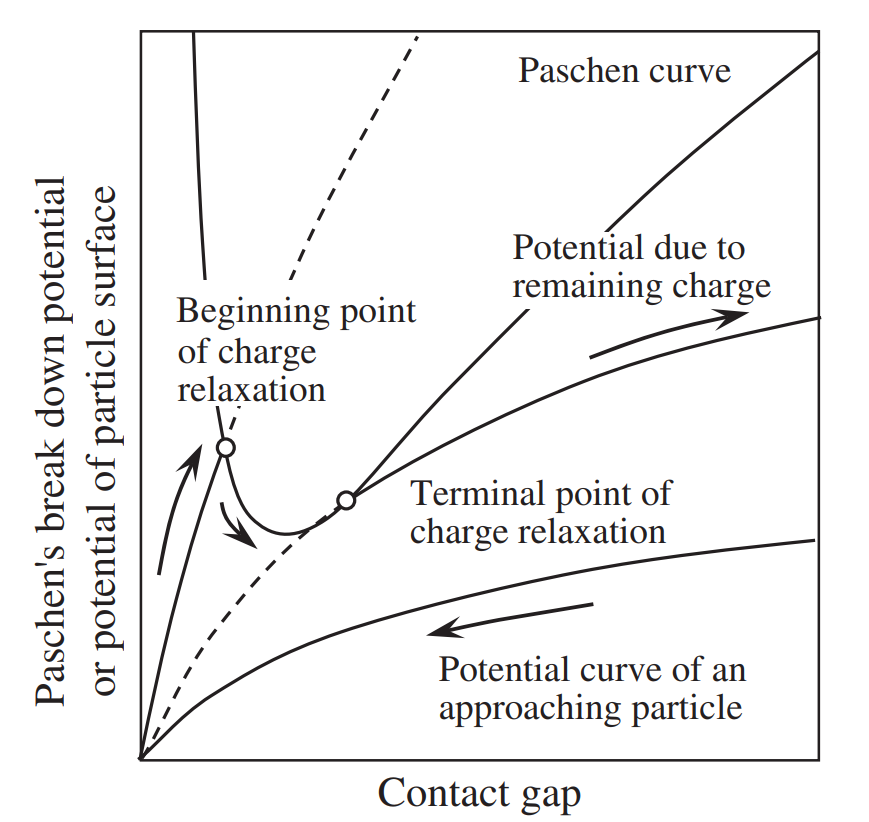}
\caption[]{\hl{Charge relaxation model \mbox{\citep{Mat95c,Mat10}}
(reprinted with permission).}}
\label{fig:Mat10}
\end{figure}

Each particle has a maximum charge it can hold, its \textit{equilibrium} or {saturation charge}.
When studying the electric field's effect on particles in pneumatic conveyors, it is sometimes assumed that the particles reached their saturation charge already during previous handling or operations~\citep{Yao20,Yan21}.
This is the worst-case scenario, where the electric field of the particles and conveyor walls affects the powder flow the strongest.

In pneumatic conveying, particles can saturate in two ways:
Either, they undergo charging during repeated contacts until they reach their electric equilibrium.
Or, the charge they hold is higher than what the surrounding gas can withstand \hl{before it ionizes due to its finite dielectric strength.
Then, the charge will relax. 
Several steps precede charge limitation by gas ionization.
The sequence of steps} is reflected by the \textit{charge relaxation model}~\citep{Mat95c}, whose principle is visualized in Fig.~\ref{fig:Mat10}.
The arrows present the evolution of the potential difference between the particle and the wall, which increases after contact.
\hl{In the beginning, a low-charged particle approaches the surface.
When the particle touches the other surface, both surfaces exchange charge.
In the given example, the amount of transferred charge is higher than what the surrounding medium can withstand.
Therefore, during retraction, the electric field exceeds the breakdown potential (here, given by Paschen's law).
The charge on the particle relaxes from its oversaturated state to a value that the surrounding medium can support. 
Thus, the relaxation mechanism is ionization; 
the ions produced during dielectric breakdown neutralize partly the particle charge.}
Therefore, this model serves as an upper limit for predicted charge exchange.

\hl{This upper charge limit can be estimated through the particle's exerted electric field.}
Under atmospheric conditions, the highest electric field air can withstand is $3\cdot 10^{6}$ V/m.
Based on this value, and utilizing Gauss law~\citep{Hen06}, \citet{Jiang94} estimated the maximal charge on a particle to
\begin{equation}
\label{eq:ChargeJianOld}
Q_\mathrm{max} = 2.64 \cdot 10^{-5} p \pi d_\mathrm{p}^{2}  
\qquad
\text{where}
\qquad
p=\dfrac{3 \, \varepsilon_\mathrm{r}}{2+\varepsilon_\mathrm{r}} \, .
\end{equation}
Therein, $p$ ranges from~3 to unity for a relative permittivity of $\infty \geq \varepsilon_\mathrm{r} \geq 1$.

Based on their experimental data, \citet{Mat10} estimated 
\begin{equation}
\label{eq:ChargeTatSole}
Q_\mathrm{max} = 6.43 \cdot10^{-6} d_\mathrm{p}^{3/2} \, .
\end{equation}
Due to the exponent of particle diameter, Eq.~(\ref{eq:ChargeTatSole}) predicts higher saturation charges for small particles than Eq.~(\ref{eq:ChargeJianOld}).

\citet{Mat10a} extended the charge relaxation model for the effects of the surrounding charged particles and image charges.
They defined two regions:
In the region where neither surrounding particles nor image charge play a role, the maximum charge is
\begin{equation}
\label{eq:ChargeRelaxLimit}
Q_\mathrm{max} = \dfrac{2 \pi \varepsilon}{3 \xi D} E_\mathrm{crit} d_\mathrm{p}^{3} \, ,
\end{equation}
where $D$ is the diameter of the transport pipe, $E_\mathrm{crit}$ the critical electric field, and $\xi$ the solid volume fraction.
In the region where both effects reduce the maximal charge, combining Eqs.~(\ref{eq:ChargeTatSole}) and (\ref{eq:ChargeRelaxLimit}) yields an empirical equation valid in both regions, namely 
\begin{equation}
\label{eq:ChargeRelaxLimitEmpiric}
Q_\mathrm{max}= \dfrac{1.1\cdot10^{-4} d_\mathrm{p}^{3}}{\sqrt{(\xi D)^2 + (17.1 \, d_\mathrm{p}^{1.5})^2}}
\end{equation}

Although accurately estimated charge limits are important, generally particles carry different charge.
If particles collide frequently with each other, they can even carry charge of different polarity.
Particles with different polarities interact completely different compared to unipolar particles.

Moreover, from a safety perspective the particles' charging dynamics might even be more important than their maximum charge.
A particle contacting a surface separates charge between both objects.
While the charge build-up of the particle is limited to $Q_\mathrm{max}$, the charge accumulation on the plants component might be incomparably larger.
Then, the danger to the facility stems the charging dynamics, i.e., the rate at which charge transfers to the component.

\subsection{Condenser models}

The most spread CFD model to predict particle contact charging dynamics is the so-called \textit{condenser model} that has been in use for more than five decades~\citep{Soo71,Mas76,John80}.
Its name refers to the analogy of particle charging to the temporal response of a capacitor (also known as a condenser) in a resistor-capacitor (R-C) circuit.
Even though the condenser model appeared over the years in different variants, all formulations are based on the same assumptions:
\begin{enumerate}
\item A particle charges upon contact with another surface.
\item The driving force for the charge transfer is the contact potential difference of the material pair, $V$, and the charge held by the particle before contact.
\item The direction of transferred charge is driven by the $V$, which might be a sum of multiple contributing factors.
\item The amount of transferred charge depends on the material properties and the contact kinematics.
\item The particle charge saturates asymptotically.
\end{enumerate}

Thus, during collisions of two particles of the same material, which is the typical situation for particles being part of the same powder batch, no charge transfers because their contact potential is the same.
Nevertheless, charge may exchange if at least one of the two particles carries a charge prior to the contact.
In the original formulation by \citet{Soo71}, the charge transfer between two particles, $\Delta Q_n = -\Delta Q_m$, during the collision contact time, $\Delta t_{\mathrm{p}}$, reads
\begin{equation}
\label{eq:chargeexpp}
\Delta Q_n = \dfrac{C_n C_m}{C_n+C_m} \left( \dfrac{Q_m}{C_m} - \dfrac{Q_n}{C_n} \right) \left( 1- \mathrm{e}^{-\Delta t_{\mathrm{p}} / \tau_{\mathrm{p}}} \right) = -\Delta Q_m \, .
\end{equation}
In the above equation, $C_n$ and $C_m$ denote the capacity of both particles and $\tau_{\mathrm{p}}$ their charge relaxation time.

Afterward, \citet{John80} expanded the model to the impact of a spherical particle with a plane surface such as a wall or a plate.
In opposite to particle-particle collisions, in this situation, the two objects in contact are usually of dissimilar material.
For the contact potential to be the driving force, at least one contacting material needs to be a conductor.
This assumption is met if the conveyor walls are made of metal.
The condenser model appeared in two different forms.
The first substitutes the contact potential with the contact capacitance,
\begin{equation}
\label{eq:chargeResitance}
\frac {dQ}{dt_c} = \frac{1}{R_0} (V_0-V_\mathrm{c}) = \frac{A_{\mathrm{pw}}}{R_0 C} (\sigma_0-\sigma) =\frac{A_{\mathrm{pw}}}{\tau} (\sigma_0-\sigma) \, .
\end{equation}
Herein, $t_c$ is the charging duration.
This approach uses the experimentally obtained equilibrium charge density, $\sigma_0$, and the characteristic time-scale $\tau=R_0 C$ as fitting parameters.
However, this equation is only valid if a particle collides with a conductive wall because $\sigma_0$ implicitly includes the contact potential.
Analogous, but in terms of the number of successive impacts, $n$, instead of the charging time, the model reads
\begin{equation}
\label{eq:CondeserTat}
\frac{dQ}{dn} = k_\mathrm{c} (V_\mathrm{c} - k_0 Q)  -\frac{k_\mathrm{r}}{f} Q \, .
\end{equation}

\begin{figure} [tb]
\centering
\includegraphics[trim=0cm 6.4cm 0cm 0cm,clip=true,width=0.85\textwidth] {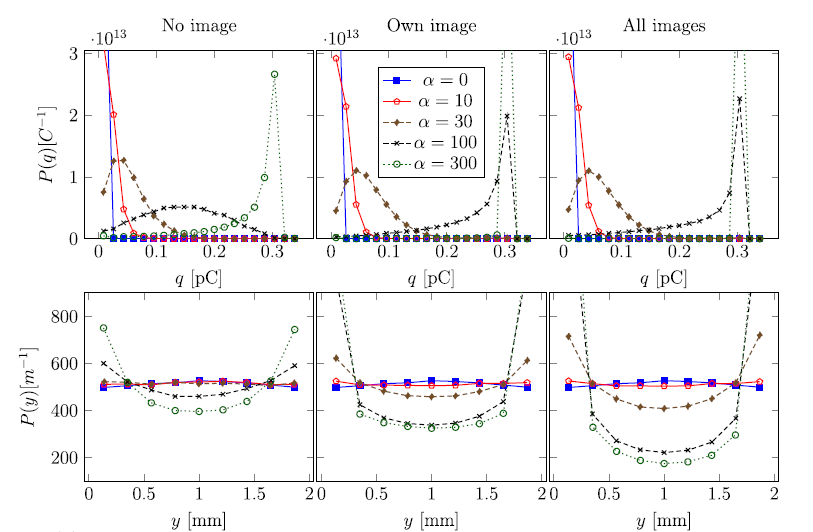}
\caption{
The particles' charge pdf in a duct flow. 
The charging model considers no image charges (left), only the particles' own image charge (middle), or all image charges (right).
The parameter $\alpha$ denotes the charging efficiency, which enters Eq.~(\ref{CondModelSimple}) through $c=\alpha A_\mathrm{c}$~\citep{Kor14}
(reprinted with permission).}
\label{fig:ImageCharge}
\end{figure}

The second, more generalized form, of the condenser model is
\begin{equation}
\label{CondModelSimple}
\Delta Q = c \left(1-\frac{q_\mathrm{in}}{q_\mathrm{sat}} \right) \, .
\end{equation}
Herein, $c$ includes various parameters affecting charging, for example slip or effective contact area.
\citet{Kor14} defined $c=\alpha A_\mathrm{c}$, where $A_\mathrm{c}$ is the contact area and $\alpha$ is an empirical parameter.
In their model, $c$ estimates the probability of collision with an uncharged patch on another particle's surface.
Figure~\ref{fig:ImageCharge} depicts the effect of $\alpha$ on the particles' charge distribution;
the higher $\alpha$ the more \hl{particles} reach their saturation charge during \hl{transport} in a pipe.
Moreover, the figure demonstrates the tremendous influence of image charges on the resulting charge distributions.
Other versions of this model describe the charging of insulator particles~\citep{Pei13}.

All above discussed models can be equivalently written in integral form.
For example, integrating Eq.~(\ref{eq:CondeserTat}) with the initial conditions and $Q=Q_0$ and $k_\mathrm{r}=0 $ and assuming that all parameters are constant yields 
\begin{equation}
\label{ExpChargingPW}
Q= Q_0 \exp \left(-\frac{n}{n_0} \right)+Q_\infty \left(1- \exp \left(-\frac{n}{n_0}\right) \right) \, ,
\end{equation}
where $Q_\infty$ is the saturation charge, $n_0$ the characteristic number of collisions, and $Q_0$ the initial charge.
In the above equation, the charge increases exponentially during successive contacts;
a characteristic form of condenser-based models.

As mentioned above, the condenser model went through some evolutionary steps, one being the refinement of the contact potential difference to~\citep{Mat00} 
\begin{equation}
\label{VoltageRefined}
V = V_\mathrm{c} - V_\mathrm{im} -  V_\mathrm{b} + V_\mathrm{ex} \, .
\end{equation}
Therein, the $V_\mathrm{c}$ is the contact potential derived from the materials' work functions.
The effect of image charges, $V_\mathrm{im}$, becomes significant if a charged particle impacts a conductive surface where it induces a charge of opposite polarity and, thus, increases the driving force.
The contribution of space charge to the contact potential, $V_\mathrm{b}$, reflects other charged particles in the vicinity.
The electric field created by surrounding charged particles reduces the saturation charge because it de-localizes the induced image charge.
Finally, external electric fields might affect the contact potential by $V_\mathrm{ex}$.

The refinement of $V$ allows for the modeling of particle/particle interactions~\citep{Chow21c,Han21}.
The precise calculation of the electric field at their contact point yields the charge diffusion between two particles of the same material but a different charge, their charge exchange due to polarization (\textit{dipole-amplification model}), 
and the charging of multiple component mixtures.
Also, this approach expedites recent Eulerian formulation of powder charging~\citep{Ray20,Ray18,Kol18}.

\begin{figure}[tb]
\centering
\subfloat[]{\includegraphics[trim=0mm -15mm 0mm 0mm,clip=true,width=.22\textwidth]{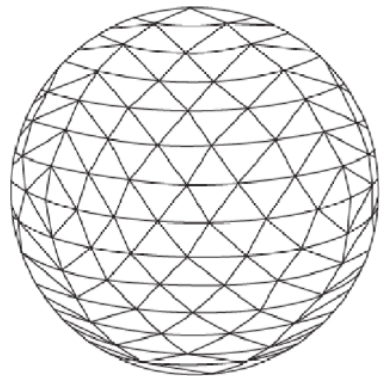}%
\label{fig:Yos03}%
}
\qquad
\subfloat[]{
\includegraphics[trim=0cm 0cm 0cm 0cm,clip=true,width=.62\textwidth]{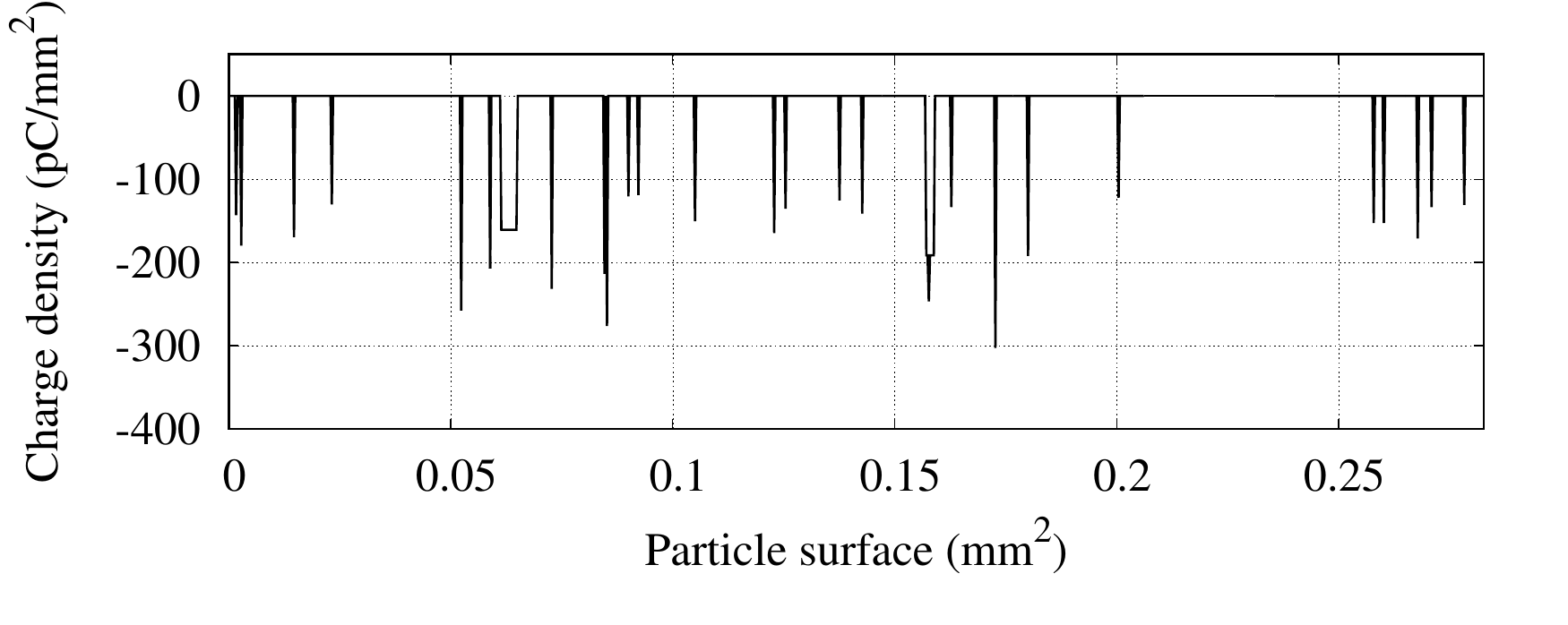}
\label{fig:Gro16f}}
\caption{(a) Charging site concept of \citet{Yos03}.
(b) Resolved charge density on a particle's surface after pneumatic conveying \citep{Gro16f}
(reprinted with permission).}
\end{figure}

The above formulations of the condenser model assume a uniform charge distribution on the particle's surface.
However, the charge does not distribute uniformly on insulative surfaces, such as polymers.
Therefore, the strong scattering of the impact charge in the single-particle experiments of \citet{Mat03} was attributed to a non-uniform charge distribution on the particle's surface.
As a response to the observed scatter, several models resolve the charge location on particle surfaces.
\citet{Yos03} introduced the concept of charging sites that take up charge individually, see Fig.~\ref{fig:Yos03}.
Using the concept of charging sites, \citet{Gro16f} extended the condenser model to the \textit{non-uniform charge model} for particle/surface and inter-particle collisions.
This formulation leads to a wide range of possible outcomes of contact, which partly explains the scatter of the measured charging of single PTFE particles.
The non-uniform charge model was used to simulate pneumatic powder transport.
Figure~\ref{fig:Gro16f} shows the resolved charge on the particle surface after leaving the duct.
Each impact causes one peak.
Some peaks overlap, which means the particle impacted at a location where a previous impact already left a charge spot.

Because of its simplicity, the condenser model's serves as basis for most new models that aim to include different contact modes.
Those extensions treat bouncing or sliding or non-spherical particles~\citep{Ire08,Ire10,Ire12,Ire19}.
Rolling during contact was included in the LES of~\citet{Li21}.

Further, the condenser model was adapted to model reduced charging in humid environments and due to space charge~\citep{Kol17}.
\hl{Also, changing the effective work function led to size-dependent charging of particles of the same material~\mbox{\citep{Liu20}}.
Another attempt to model same-material charging with the condenser model is to reflect variations in the local chemical compositions by spatially adapting the work function~\mbox{\citep{Xucheng19}}.
Nevertheless, surface state models (see next section) seem better suited to model charge transfer at the nanoscale between surfaces of the same material.
}

In general, condenser models fit experimental data if their model parameters, e.g., the effective work function, are tuned carefully.
However, the condenser model remains a phenomenological approach because it fails to provide any information that go beyond the charging of conductors upon simple contact.

\subsection{Surface states models}

Another group of charging models relies on the surface state theory~\citep{Low86a,Low86b}.
According to it, excited electrons exist on the surface of insulators.
They are trapped and immobile because of the insulator's high surface resistivity, even though the insulator surface has vacant states.
The only way to relax is when another surface with vacant states approaches these excited electrons; 
then, charge transfer occurs.
Surface states models assume the contact potential to be much weaker than local inhomogeneities of the effective work functions due to excited states.

These models aim to explain the charging of particles made of the same material.
The low-density limit was recently utilized in models~\citep{Duff08}, in a probabilistic version~\citep{Lacks07}, and in a more general formulation considering the transfer of any charged species~\citep{Kon17}.
The low-density limit assumes that one concentration of one charged species is significantly lower than of the other one.
By assuming the transfer of charge carriers from one particle to another until they are depleted, the results of this model agreed with two trends observed in powder flows:
particles charge stronger in highly poly-disperse systems, and big particles are usually positively and small particles negatively charged.

Recent experimental~\hl{\mbox{\citep{Som20,Xucheng19,Bai21}}} and numerical~\hl{\mbox{\citep{Lacks19,Mizzi19,Grosj20}}} progress point toward surface states models for particle/particle charging, supported by a rigorous description of the underlying charging mechanisms.
However, a definite model and suitable models for CFD simulations remain outstanding.

\subsection{Empirical models}

\begin{figure}[tb]
\vspace{-\ht\strutbox}
\centering
\includegraphics[trim=0cm 0cm 0cm 0cm,clip=true,width=.47\textwidth]{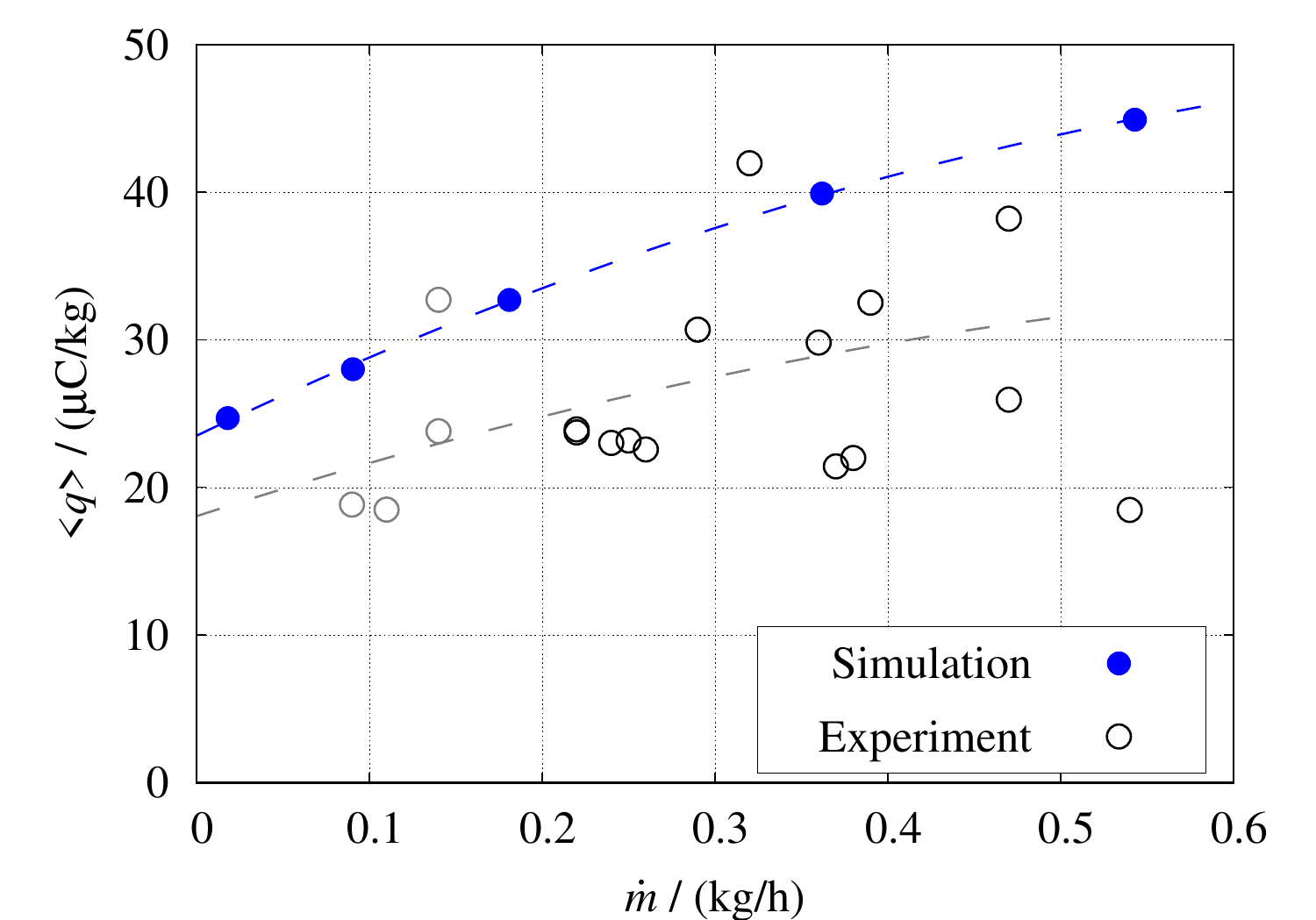}
\caption{CFD simulations of the electrification of PMMA particles during conveying using an empirical charging model.}
\label{fig:Gro21c}
\end{figure}

Finally, a purely empirical charging model was recently proposed by \citet{Gro21c} for spherical PMMA particles.
The model bases on data from single-particle experiments using the precise same particles as in the simulations.
The CFD simulations agree well with experiments, see Fig.~\ref{fig:Gro21c}, for 200~$\upmu$m particles but fail for 100~$\upmu$m particles.

\hl{Another new empirical model by \mbox{\citet{Han21}} proposes to reduce the complexity of charging models by introducing measured values of the charge transfer limit and charge efficiency.
Therein, an electric field realizes the limit at which the charge transfer halts.
This implementation modifies the condenser model.
The contained empirical values aim to capture all relevant parameters that affect charging: 
humidity, impact velocity, material composition, temperature, etc.
Currently, their model is validated with single-particle experiments;
the predictions in complex systems requires further research.}

However, like all the above-discussed charging models, empirical models handle only specific situations.
A generally predictive charging model that satisfies the requirements of a CFD tool is not in reach yet.
Thus, the particle charging model will remain the largest contributor to the overall error of CFD simulations of powder flow electrification in the foreseeable future.

\section{Conclusions and perspectives for future research}

Due to its extraordinary complexity, the CFD simulation of powder electrification \hl{during pneumatic conveying} has failed so far.
It requires the solution of an interdisciplinary mathematical model describing turbulence, electrostatics, and triboelectric charging.
This paper reviewed the state-of-the-art and pinpointed the future research necessary to improve the numerical predictions.

\hl{New multi-physics and multi-scale approaches can significantly improve the simulation of powder flow charging.
At the nano-scale, current efforts aim to capture same-material charging, bipolar charging, and the effect of particle polarization.
To predict the charging of complete powders, these models need to be incorporated into computationally efficient macroscopic simulation frameworks.}
Highly resolved large-eddy and direct numerical simulations of the carrier gas flow combined with Lagrangian simulations of the particle dynamics offer insight into the detailed mechanics of powder charging.

The particle charging model is the largest contributor to the current simulation error.
\hl{Past pneumatic conveying simulations often ignored the build-up of particle charge but assumed a constant charge instead.
However, experiments and recent numerical predictions demonstrate a strong dependence of powder charging on the gas flow pattern.
These findings underline the necessity for further development of dynamic particle charging models in conjunction with CFD.}
Understanding the dependence of the powder charging rate on the conveyor operating parameters, such as velocity or powder mass flow rate, can guide the design of future safe conveying systems.

A generally valid, predictive model seems currently out of reach.
But new single-particle experiments that deliver impact data tailored to pneumatic conveying can improve the accuracy of models for specific particles.

Finally, recent Eulerian-Eulerian formulations open a way to simulate powder charging not only in a limited pipe section but in complete pneumatic conveyors.
The next step to improve Euler-Euler models is their handling of polydisperse particle size distributions.
\hl{
Also, Eulerian descriptions of electrostatically charged powder lack models for deposits.
Implementing polydispersity and deposits is straightforward in the Lagrangian but strenuous in the Eulerian framework.
The direct quadrature method of moments was recently applied to describe the evolution of polydisperse particles carrying charge.
Thus, moment-based methods seem promising.
However, further research is required to prove the viability of Eulerian approaches to simulate powder charging in complete flow facilities.
}



\section*{Acknowledgements}
This project has received funding from the European Research Council~(ERC) under the European Union’s Horizon 2020 research and innovation programme~(grant agreement No.~947606 PowFEct).

\section*{Conflict of interest}
The authors declare that there is no conflict of interest regarding the publication of this article.

\bibliography{publications}
\end{document}